\documentstyle[eqsecnum,prb,aps,twocolumn,epsbox]{revtex}
\begin{document}
\twocolumn[
\begin{center}
{\large{\bf Theory of the metamagnetic crossover
in CeRu$_2$Si$_2$}}

\vspace{5mm}
{Hiroyuki Satoh
and Fusayoshi J. Ohkawa}
\par

{\small\em Division of Physics, Hokkaido University, 
Sapporo 060-0810, Japan}
\par

%({\today})
({November 8, 2000})
\vspace{3mm}

{\small \parbox{142mm}{\hspace*{5mm}
Based on the periodic Anderson model,
it is shown that the competition between the quenching 
of magnetic moments by local quantum spin fluctuations
and a magnetic exchange interaction caused by the virtual 
exchange of pair excitations of quasiparticles in spin 
channels is responsible 
for the metamagnetic crossover in CeRu$_2$Si$_2$,
cooperated with the electron-lattice interaction.
The strength of the exchange interaction 
is proportional to the bandwidth of quasiparticles
and its sign %depends on the magnetic field;
changes with increasing magnetizations;
it is antiferromagnetic %at zero field, 
in the absence of magnetizations,
whereas
it is ferromagnetic in the metamagnetic crossover region.
Experimental results of static quantities
are well reproduced.
\vspace{2mm} \\
1999 PACS: 71.10.-w, 71.27.+a, 75.30.Et, 75.30.Kz
}
}

\end{center}

\vspace{8mm}
]

%
%
%********************************************************
\section{Introduction}
%********************************************************
%
The metamagnetic crossover in CeRu$_2$Si$_2$ is
an important issue.
The compound has a large electronic specific heat
coefficient of $\gamma \simeq 360$~mJ/mol~K$^2$,
\cite{heat}
and it shows a sharp increase of magnetization
at the field of $H_M\simeq 7.7$ T.
\cite{metamag,Sakakibara}
Other physical properties such as magnetostriction
\cite{scaling,striction}
and specific heat
\cite{gamma}
are also anomalous in this field region.
One of the most crucial experimental results 
to be explained is
the single-parameter scaling;
\cite{scaling,scaling2}
independent experimental quantities
for different pressures are scaled with
a single energy parameter $k_BT_K$, 
where $T_K$ is called the Kondo temperature.
The Kondo temperature is the energy scale of
local quantum spin fluctuations and is 
approximately equal to the bandwidth of quasiparticles.

The present authors showed in a previous paper 
\cite{pre} 
that an exchange interaction caused by the virtual 
exchange of pair excitations of quasiparticles in spin 
channels plays a critical role in the metamagnetic 
crossover as well as the electron-lattice interaction
called the Kondo volume-collapse effect.
\cite{collapse}
This exchange interaction has the following two novel
properties.
First, it
changes from being antiferromagnetic at zero fields to 
being ferromagnetic in the metamagnetic crossover region 
because of a pseudogap structure 
in the density of quasiparticle states,
which is characteristic of heavy-electron compounds.
Second, the strength of the exchange interaction 
is proportional to the bandwidth of quasiparticles.
Then, 
the single-parameter scaling can be easily explained. 

After the previous paper was submitted,
a detailed measurement of the field dependence of
specific heat was reported by Aoki {\it et al}. 
\cite{double-peak}
At sufficiently low temperatures,
a single sharp peak was observed at $H_M$,
while a double-peak structure was found at higher
temperatures. 
They argued that such a result can be explained 
if a sharp peak exists in the density of states.
A similar argument applies straightforwardly to our 
previous model; however, if the pseudogap is enough deep,
the magnetization process shows a first-order 
transition within the theoretical framework of 
the previous paper.
Recently, two theories were proposed.
\cite{Ono,Ohara}
They claimed that an anisotropic $c$-$f$ mixing
plays an important role in the metamagnetic behavior.
In both theories, however, the peak structure 
in the density of states is too sharp 
because the ${\bf k}$ dependence of hybridization 
matrices is improperly treated;
one-dimensional van Hove singularity is irrelevant.
Such an extremely sharp peak is inconsistent with 
the experimental result of specific heat.
Furthermore, it should give a discontinuous transition 
instead of the metamagnetic crossover
if the novel exchange interaction mentioned above is 
properly taken into account.
To reproduce the experimental results 
of magnetization and specific heat simultaneously
has not been achieved so far.

A main purpose of this paper is 
to reformulate and improve the previous theory 
based on the periodic Anderson model 
so as to explain static properties of the system.
We will also give a brief comment on intersite
spin fluctuations around the zone center.
%
%
%********************************************************
\section{Model}
%********************************************************
%
The periodic Anderson model is written as
\begin{eqnarray}
{\cal H} &=& 
\sum_{\lambda{\bf k}\sigma} E_{\lambda}({\bf k})
a^{\dag}_{\lambda{\bf k}\sigma}a_{\lambda{\bf k}\sigma}
+ \sum_{{\bf k}\sigma}(E_{f} - \sigma H^*)
f^{\dag}_{{\bf k}\sigma}f_{{\bf k}\sigma} \nonumber \\
&& + \sum_{\lambda{\bf k}\sigma} [ V_{\lambda}({\bf k})
a^{\dag}_{\lambda{\bf k}\sigma}f_{{\bf k}\sigma}
+ {\rm h.c.}]
+\frac1{2}U\sum_{i\sigma} n_{i\sigma}n_{i-\sigma},
\nonumber \\
\label{PAM} %----------------------------------------
\end{eqnarray}
with $\lambda$ the band index of conduction electrons,
$n_{i\sigma} = f^{\dag}_{i\sigma} f_{i\sigma}$, 
and $H^* = m_0 H$, 
where $m_0$ is the saturation magnetization
per $f$ electron.
Other notations are standard.
The kinetic energy of conduction electrons,
$E_{\lambda}({\bf k})$,
and the $f$ electron level, $E_{f}$, are measured
from the chemical potential, respectively.

When multiple conduction bands are assumed,
a pseudogap structure is expected in the density of
states because of the hybridization between the $f$ band
and conduction bands.
\cite{pseudo-gap}
Since the property of the magnetic exchange interaction
that plays a critical role in the magnetization process
depends on the shape of the density of quasiparticle 
states as shown in subsequent sections,
a phenomenological model for the density of quasiparticle 
states is employed instead of considering explicit forms
of $E_{\lambda}({\bf k})$ and $V_{\lambda}({\bf k})$
for simplicity.
Furthermore, in order to make our treatment easy,
we assume that the system is symmetrical.
In other words, the Hamiltonian~(\ref{PAM}) is assumed to 
be invariant under the particle-hole transformation.

One of the most crucial issues in constructing a theory 
of strongly correlated electron systems is how properly
local quantum spin fluctuations, which are responsible 
for the quenching of magnetic moments, are treated.
They can be correctly taken into account in the 
single-site approximation (SSA)
\cite{SSA} 
that is rigorous for paramagnetic states 
in infinite dimensions.
Consider the zero-field case at first; $H=0$.
Within the SSA, Green's functions for $f$ electrons and 
conduction electrons are respectively given by
\begin{equation}
G_{ff\sigma}(i\varepsilon_n,{\bf k}) = 
\frac1{i\varepsilon_n - E_f 
- \tilde{\Sigma}_{\sigma}(i\varepsilon_n) 
- \displaystyle{
\sum_{\lambda}\frac{|V_{\lambda}({\bf k})|^2}
{i\varepsilon_n - E_{\lambda}({\bf k})}}},
\label{PAM-Green1} %-------------------------------------
\end{equation}
with $\tilde{\Sigma}_{\sigma}$ 
the single-site self-energy function, and
\begin{eqnarray}
\lefteqn{
G_{\lambda\lambda'\sigma}(i\varepsilon_n,{\bf k}) =
\delta_{\lambda\lambda'}
g_{\lambda}(i\varepsilon_n,{\bf k})} \nonumber \\
&&
+ g_{\lambda}(i\varepsilon_n,{\bf k})
V_{\lambda}({\bf k})
G_{ff\sigma}(i\varepsilon_n,{\bf k})V^*_{\lambda'}({\bf k})
g_{\lambda'}(i\varepsilon_n,{\bf k}),
\label{PAM-Green2} %-------------------------------------
\end{eqnarray}
with
$g_{\lambda}(i\varepsilon_n,{\bf k}) =
[i\varepsilon_n - E_{\lambda}({\bf k})]^{-1}$.
Here, $i\varepsilon_n$ is an imaginary fermion energy 
with $n$ an integer.
To obtain the single-site self-energy function
is reduced to solving a single-impurity Anderson model
\cite{SSA}
that has the same localized electron level $E_f$ and 
on-site Coulomb repulsion $U$ as those in Eq.~(\ref{PAM}). 
Call this Anderson model a mapped Anderson model (MAM).
Other parameters in the MAM are determined through the
mapping condition:
\begin{equation}
\tilde{G}_{ff\sigma}(i\varepsilon_n) =
\frac1{N}\sum_{\bf k}G_{ff\sigma}(i\varepsilon_n,{\bf k}),
\label{mapping-condition} %-----------------------------
\end{equation}
where $N$ is the number of unit cells and
$\tilde{G}_{ff\sigma}$ is the Green's function
of the MAM written as
\begin{equation}
  \tilde{G}_{ff\sigma}(i\varepsilon_n) =
  \frac1{i\varepsilon_n - E_f
   - \tilde{\Sigma}_{\sigma}(i\varepsilon_n)
   - L(i\varepsilon_n)},
\label{MAM-Green} %------------------------------------
\end{equation}
with
$L(i\varepsilon_n) = (1/\pi)
\int_{-\infty}^{\infty} d\varepsilon
\ \Delta(\varepsilon)/(i\varepsilon_n - \varepsilon)$.
Here, $\Delta(\varepsilon)$ is the hybridization energy
of the MAM.
Once a trial function for $\Delta(\varepsilon)$ is given,
$\tilde{\Sigma}_{\sigma}(i\varepsilon_n)$ is obtained
by solving the MAM numerically.
Therefore, Eq.~(\ref{mapping-condition}) is a
self-consistency condition for $\Delta(\varepsilon)$.
However, we do not perform this self-consistent
calculation in this paper.
Instead, several approximations will be taken 
in subsequent sections
with the use of well-known results for the Kondo problem.

Consider the spin susceptibility of the MAM,
$\tilde{\chi}_s(i\omega_l)$,
where $i\omega_l$ is a boson energy with $l$ an integer.
The Kondo temperature for the periodic Anderson 
model, which is the energy scale of 
local quantum spin fluctuations, is defined by
\begin{equation}
\lim_{T \to 0}\tilde{\chi}_s(+i0)
\equiv
\frac1{k_BT_K}.
\label{TK-def} %---------------------------------
\end{equation}
In the same way as the previous paper, 
the electron-lattice interaction is taken into account 
simply through the volume dependence of the Kondo 
temperature:
\begin{equation}
T_K(x) = T_K(0)e^{-x},
\label{TK-volume} %---------------------------------
\end{equation}
with $x = \Gamma\ \Delta V/V_0$.
Here, $\Gamma \sim 190$ is the Gr\"{u}neisen constant
of $T_K$. 
For the sake of simplicity, the argument $x$ will be 
omitted unless particularly required hereafter.
%
%
%********************************************************
\section{Fermi liquid description}
%********************************************************
%
The self-energy function is expanded as
\begin{equation}
\tilde{\Sigma}_{\sigma}(i\varepsilon_n) = 
\tilde{\Sigma}_{\sigma}(+i0)
+[1 - \tilde{\phi}_m]i\varepsilon_n + \cdots,
\end{equation}
for small $|\varepsilon_n|$, where $\tilde{\phi}_m$ 
is a mass enhancement factor in the SSA.
Note that
$\tilde{\Sigma}_{\sigma}(+i0) = - E_f$
in the symmetrical case.
Then the coherent part of Eq.~(\ref{PAM-Green1}) 
is written as
\begin{equation}
G^{(c)}_{ff\sigma}(i\varepsilon_n,{\bf k}) = 
\frac1{\tilde{\phi}_m i\varepsilon_n 
- \displaystyle{
\sum_{\lambda}\frac{|V_{\lambda}({\bf k})|^2}
{i\varepsilon_n - E_{\lambda}({\bf k})}}},
\label{PAM-Green3} %-------------------------------------
\end{equation}
and correspondingly, 
$G^{(c)}_{\lambda\lambda'\sigma}$ is given by 
Eq.~(\ref{PAM-Green2}) with replacing
$G_{ff\sigma}$ by $G^{(c)}_{ff\sigma}$.
Quasiparticles are defined as the poles of 
Eq.~(\ref{PAM-Green3}), 
namely, the dispersion relation of quasiparticles is
obtained by solving the following equation
\begin{equation}
\tilde{\phi}_m z 
- \sum_{\lambda} \frac{|V_{\lambda}({\bf k})|^2}
{z - E_{\lambda}({\bf k})} = 0.
\label{eq-pole1} %-------------------------------------
\end{equation}
We write the solutions as $z = \xi_{\nu}({\bf k})$
with $\nu$ representing the branch of quasiparticles.
By dividing the right-hand side of Eq.~(\ref{PAM-Green3})
into partial fractions, it follows that
\begin{equation}
G^{(c)}_{ff\sigma}(i\varepsilon_n,{\bf k}) = 
\sum_{\nu}\frac1{Z^f_{\nu}({\bf k})}\cdot
\frac1{i\varepsilon_n - \xi_{\nu}({\bf k})},
\label{PAM-Green4} %-------------------------------------
\end{equation}
where the renormalization factor is given by
\begin{equation}
\frac1{Z^f_{\nu}({\bf k})} = 
\frac1{\tilde{\phi}_m + \displaystyle{\sum_{\lambda}}
 \eta_{\lambda\nu}({\bf k})},
\label{PAM-Z1} %-------------------------------------
\end{equation}
with
$\eta_{\lambda\nu}({\bf k}) =
|V_{\lambda}({\bf k})|^2/
[\xi_{\nu}({\bf k}) - E_{\lambda}({\bf k})]^2$.
Similarly, we have
\begin{equation}
G^{(c)}_{\lambda\lambda\sigma}(i\varepsilon_n,{\bf k}) = 
\sum_{\nu}\frac1{Z^{\lambda}_{\nu}({\bf k})}\cdot
\frac1{i\varepsilon_n - \xi_{\nu}({\bf k})},
\label{PAM-Green5} %-------------------------------------
\end{equation}
with
\begin{equation}
\frac1{Z^{\lambda}_{\nu}({\bf k})} = 
\frac{\eta_{\lambda\nu}({\bf k})}
{\tilde{\phi}_m + \displaystyle{\sum_{\lambda}}
\eta_{\lambda\nu}({\bf k})}.
\label{PAM-Z2} %-------------------------------------
\end{equation}
It immediately follows from Eqs.~(\ref{PAM-Z1}) and 
(\ref{PAM-Z2}) that the renormalization factors satisfy
the relation
\begin{equation}
\frac{\tilde{\phi}_m}{Z^f_{\nu}({\bf k})}
+ \sum_{\lambda}\frac1{Z^{\lambda}_{\nu}({\bf k})}
= 1.
\label{PAM-Z3} %-------------------------------------
\end{equation}
With the use of Eq.~(\ref{PAM-Z3}), the density of 
quasiparticle states can be written in the form
\begin{eqnarray}
\lefteqn{
\rho^*(\varepsilon) \equiv
\frac1{N}\sum_{\nu{\bf k}}
\delta(\varepsilon - \xi_{\nu}({\bf k}))} \nonumber \\
&& 
= - \frac1{\pi N}\sum_{\bf k}
\ {\rm Im}\ \Biggl\{ \tilde{\phi}_m
G^{(c)}_{ff\sigma}(\varepsilon_+,{\bf k}) 
+ \sum_{\lambda}
G^{(c)}_{\lambda\lambda\sigma}(\varepsilon_+,{\bf k})
\Biggr\},
\nonumber \\
\label{PAM-rho} %---------------------------------------
\end{eqnarray}
with $\varepsilon_+ = \varepsilon + i0$.
The energy scale for quasiparticles, $T_Q$, is defined by
\begin{equation}
\rho^*(0) \equiv \frac1{4k_BT_Q}.
\label{TQ-def} %---------------------------------------
\end{equation}
On the other hand, $Z^f_{\nu}({\bf k})$
obeys the following condition
\begin{equation}
\sum_{\nu}\frac1{Z^f_{\nu}({\bf k})} = 
\frac1{\tilde{\phi}_m},
\label{PAM-Z4} %-------------------------------------
\end{equation}
which can be proved by comparing Eqs.~(\ref{PAM-Green3})
and (\ref{PAM-Green4}) for the limiting case of 
$|\varepsilon_n| \to + \infty$. 
It should be noted that 
$Z^f_{\nu}({\bf k}) \simeq \tilde{\phi}_m$ 
for the quasiparticle band 
that is closest to the Fermi level and
$Z^f_{\nu}({\bf k}) \gg \tilde{\phi}_m$ 
for other branches for a given ${\bf k}$, that is,
\begin{equation}
\frac{\tilde{\phi}_m}{Z^f_{\nu}({\bf k})}
\simeq \left\{
\begin{array}{cc}
1 & \mbox{\hspace{5mm}for\hspace{5mm}} 
|\xi_{\nu}({\bf k})| \alt 2k_BT_Q \\ \\
0 & \mbox{\hspace{5mm}for\hspace{5mm}} 
|\xi_{\nu}({\bf k})| \gg 2k_BT_Q
\end{array}
\right..
\label{phi/Z} %--------------------------------------
\end{equation}
Consider a spectral function of 
renormalized $f$ electrons, $\rho^*_{f}$, defined by
\begin{equation}
\frac1{N}\sum_{\bf k}
G^{(c)}_{ff\sigma}(i\varepsilon_n,{\bf k})
= \frac1{\tilde{\phi}_m}
\int_{-\infty}^{\infty}d\varepsilon
\ \frac{\rho^*_{f}(\varepsilon)}
{i\varepsilon_n - \varepsilon},
\label{PAM-spectral-1} %-------------------------------
\end{equation}
or equivalently
\begin{equation}
\rho^*_{f}(\varepsilon) =
\frac1{N}\sum_{\nu{\bf k}}
\frac{\tilde{\phi}_m}{Z^{f}_{\nu}({\bf k})}
\ \delta(\varepsilon - \xi_{\nu}({\bf k})).
\label{PAM-spectral-2} %-------------------------------
\end{equation}
Equation~(\ref{phi/Z}) tells that
\begin{equation}
\rho^*_{f}(\varepsilon)
\simeq \left\{
\begin{array}{cc}
\rho^*(\varepsilon) & \mbox{\hspace{5mm}for\hspace{5mm}} 
|\varepsilon| \alt 2k_BT_Q \\ \\
0 & \mbox{\hspace{5mm}for\hspace{5mm}} 
|\varepsilon| \gg 2k_BT_Q
\end{array}
\right.,
\label{PAM-spectral-3} %-------------------------------
\end{equation}
and it follows from Eq.~(\ref{PAM-Z4}) that
$\rho^*_{f}$ satisfies the normalization condition of
$\int_{-\infty}^{\infty}d\varepsilon\ 
\rho^*_{f}(\varepsilon) = 1$.

The Luttinger's argument 
\cite{Luttinger}
applies straightforwardly to the periodic Anderson model.
Consider the number of electrons per unit cell:
\begin{equation}
n_{\sigma} = 
\frac1{N\beta}\sum_{n{\bf k}}e^{+i\varepsilon_n0}
\left\{ G_{ff\sigma}(i\varepsilon_n,{\bf k})
+ \sum_{\lambda}
G_{\lambda\lambda\sigma}(i\varepsilon_n,{\bf k}) \right\},
\label{PAM-num} %---------------------------------------
\end{equation}
with $\beta = 1/k_BT$.
At $T=0$ K, Eq.~(\ref{PAM-num}) can be transformed to
\begin{eqnarray}
n_{\sigma} =
\int_{-\infty}^{0}d\varepsilon\ \rho^*(\varepsilon),
\label{PAM-num2} %---------------------------------------
\end{eqnarray}
which is an exact relation and is identical to
the Luttinger's theorem.
Let us derive an expression for the specific heat.
Recently an exact expression for the entropy of 
an interacting system was established.
\cite{Kita}
For the periodic Anderson model, it is written as
\begin{eqnarray}
\lefteqn{
S =
-\frac1{\pi N}\sum_{{\bf k}\sigma}\int_{-\infty}^{\infty}
d\varepsilon\ \frac{\partial f(\varepsilon)}{\partial T}
\Biggl\{
\sum_{\lambda}
{\rm Im}
\ \ln\left[- g^{-1}_{\lambda\sigma}(\varepsilon_+,{\bf k}) 
\right]} \nonumber \\
&& 
+{\rm Im}
\ \ln\left[ - G^{-1}_{ff\sigma}(\varepsilon_+,{\bf k}) 
\right]
+{\rm Re}\ G_{ff\sigma}(\varepsilon_+,{\bf k})
\cdot {\rm Im}
\ \tilde{\Sigma}_{\sigma}(\varepsilon_+) \Biggl\},
\nonumber \\
\label{exact-entropy} %--------------------------------
\end{eqnarray}
where $f(\varepsilon) = 1/( e^{\beta\varepsilon}+1)$.
In Eq.~(\ref{exact-entropy}), $G_{ff\sigma}$ can be 
replaced by its coherent part because
the derivative of $f(\varepsilon)$ is non-zero only for
small $|\varepsilon|$ at low temperatures.
In differentiating the entropy with respect to $T$,
we introduce two assumptions;
${\rm Im}\ \tilde{\Sigma}_{\sigma}(\varepsilon_+) = 0$,
and
$\tilde{\phi}_m = \mbox{$T$-independent}$,
that is, quasiparticles have an infinite lifetime
and temperature-independent mass.
Since the chemical potential does not change 
as a function of temperature in a symmetrical model,
the specific heat at constant volume is given by
\begin{equation}
C_V 
= T \frac{\partial S}{\partial T}
= \sum_{\sigma} 
\int_{-\infty}^{\infty} d\varepsilon\ \varepsilon\ 
\frac{\partial f(\varepsilon)}{\partial T}
\rho^*(\varepsilon),
\label{PAM-heat} %------------------------------------
\end{equation}
where the identity
$T \partial^2 f/\partial T^2
= - \partial/\partial\varepsilon
\left( \varepsilon \partial f/\partial T \right)$
has been made use of.

The zero-temperature limit of Eq.~(\ref{PAM-heat})
is given by $C_{V} = \gamma T$ with
$\gamma =(2\pi^2k_B^2/3)\rho^*(0)$.
On the other hand,
a combination of Eqs.~(\ref{mapping-condition}),
(\ref{PAM-spectral-1}) and (\ref{PAM-spectral-3}) gives
$\rho^*(0) \simeq \rho^*_{f}(0) 
= \tilde{\phi}_m/\pi\Delta(0)$.
When $\Delta(\varepsilon)$ is constant, one can prove that
\cite{pre} 
\begin{equation}
%\tilde{\phi}_m/\pi\Delta = 1/4k_BT_K,
\frac{\tilde{\phi}_m}{\pi\Delta} = \frac1{4k_BT_K},
\label{phi-TK} %------------------------------------
\end{equation}
and then it follows $T_K \simeq T_Q$ in the SSA,
where $T_K$ and $T_Q$ are defined by 
Eqs.(\ref{TK-def}) and (\ref{TQ-def}), respectively.
We take this approximation 
in order to estimate the Kondo temperature.
From the experimental value of $\gamma$, 
we have $T_K(0) \simeq 38$ K.

Let us consider the finite-field case; $H \ne 0$.
In the previous paper,
\cite{pre}
we constructed a perturbation method
to derive a microscopic Landau free energy,
one of whose independent variables is the magnetization
$m = \sum_{\sigma} \sigma \langle
f^{\dag}_{i\sigma} f_{i\sigma} \rangle$.
In the presence of magnetizations,
physical quantities,
such as self-energy and polarization functions,
can be expressed as a function of $m$
instead of the magnetic field $H$.
The equation that defines quasiparticles becomes
\begin{equation}
\tilde{\phi}_m z 
- \delta\tilde{\Sigma}_{\sigma}(m)
- \sum_{\lambda} \frac{|V_{\lambda}({\bf k})|^2}
{z - E_{\lambda}({\bf k})} = 0,
\label{eq-pole2} %-------------------------------------
\end{equation}
with $\delta\tilde{\Sigma}_{\sigma}(m)$ a magnetic part
of the self-energy.
In Eq.~(\ref{eq-pole2}), $m$ dependence of the
mass enhancement factor has been ignored.
\cite{comment1}
The solutions for Eq.~(\ref{eq-pole2}) are denoted by
$\xi_{\nu\sigma}({\bf k},m)$.
All arguments for the finite-field case can be developed
in parallel with those for the zero-field case
with replacing 
$\xi_{\nu}({\bf k})$ by $\xi_{\nu\sigma}({\bf k},m)$. 
For example, 
$\tilde{\phi}_m/Z^f_{\nu\sigma}({\bf k},m)$
satisfies a similar property as Eq.~(\ref{phi/Z}).
Therefore, it follows by comparing 
Eqs.~(\ref{eq-pole1}) and (\ref{eq-pole2}) that
\begin{equation}
\xi_{\nu\sigma}({\bf k},m) \simeq
\xi_{\nu}({\bf k}) - \sigma\Delta E(m),
\label{Delta-E} %-----------------------------------
\end{equation}
where 
$\sigma\Delta E(m) \equiv 
- \delta\tilde{\Sigma}_{\sigma}(m)/\tilde{\phi}_m$.
With the use of Eq.~(\ref{Delta-E}),
the Luttinger's theorem gives
\begin{equation}
m = \sum_{\sigma}\sigma
\int_{-\infty}^0 d\varepsilon
\ \rho^*_{f}(\varepsilon+\sigma\Delta E(m)),
\label{sumrule1} %-------------------------------------
\end{equation}
where the polarization of conduction electrons
has been ignored.
In Eq.~(\ref{sumrule1}),
we have replaced $\rho^*$ with $\rho^*_f$;
because the right-hand side of 
Eq.~(\ref{sumrule1}) is a difference between  
the contributions from up and down spin directions, 
only the low-energy part is relevant.
From Eq.~(\ref{sumrule1}),
the magnetic part of the self-energy, $\Delta E(m)$,
can be determined as a function of $m$. 

A similar treatment for the MAM is also possible.
Consider the coherent part of Eq.~(\ref{MAM-Green}),
which is defined by
$\tilde{G}^{(c)}_{ff\sigma}(i\varepsilon_n) 
\equiv (1/N)\sum_{\bf k}
G^{(c)}_{ff\sigma}(i\varepsilon_n,{\bf k})$
and written as
$\tilde{G}^{(c)}_{ff\sigma}(i\varepsilon_n) 
=[\tilde{\phi}_m i\varepsilon_n 
- L^{(c)}(i\varepsilon_n)]^{-1}$
at zero fields. 
In the presence of magnetizations, it becomes
\begin{equation}
\tilde{G}_{ff\sigma}^{(c)}(i\varepsilon_n,m)
=
\frac1{
\tilde{\phi}_m[ i\varepsilon_n + \sigma \Delta E_A (m) ]
- L^{(c)}(i\varepsilon_n) }.
\label{MAM-Green2} %-------------------------------------
\end{equation}
Note that $\Delta E_A(m)$ is different from $\Delta E(m)$
because the MAM is determined in the absence of
magnetizations by Eq~(\ref{mapping-condition})
and then magnetic one-body potentials are added to both
the periodic Anderson model and MAM separately.
\cite{pre}
In other words, $\Delta E_A(m)$ is a single-site
term even with respect to $m$, whereas $\Delta E(m)$
includes multi-site magnetizations.
Because the magnetization in the MAM is also given by $m$
to leading order in $k_BT_K/U$,
\cite{pre} 
$\Delta E_A(m)$ can be approximately evaluated from
\begin{equation}
m = \sum_{\sigma}\sigma
\int_{-\infty}^{0} d\varepsilon\
\left( -\frac1{\pi} {\rm Im} \right)
\tilde{\phi}_m
\ \tilde{G}_{ff\sigma}^{(c)}(\varepsilon_+,m).
\label{sumrule2} %------------------------------------
\end{equation}
In case of a constant hybridization energy,
Eq.(\ref{sumrule2}) is a rigorous relation 
and is nothing but the Friedel sum rule.
It should be noted that
the evaluation of $\Delta E$ and $\Delta E_A$
from Eqs.~(\ref{sumrule1}) and (\ref{sumrule2})
is one of the most essential improvements.
In the previous paper, we assumed that
$\Delta E (m) =\Delta  E_A(m)
= (4k_BT_K/\pi)\tan (\pi m/2)$.

Before closing this section,
it is helpful to mention the volume dependence of 
the physical quantities that have appeared above.
First, it follows from 
Eqs.~(\ref{TK-volume}) and (\ref{phi-TK}) that
$\tilde{\phi}_m \propto e^{x}$.
Taking notice of Eq.~(\ref{phi/Z}),
we have 
$\rho^*_f(\varepsilon;x) 
= e^x \rho^*_f(\varepsilon e^x;0)$
and
$L^{(c)}(i\varepsilon_n;x) 
= L^{(c)}(i\varepsilon_n e^x;0)$.
Therefore, Eqs.~(\ref{sumrule1}) and (\ref{sumrule2}) 
give $\Delta E(m) \propto e^{-x}$ and
$\Delta E_A(m) \propto e^{-x}$, respectively.
%
%
%**************************************************
\section{Magnetic exchange interaction}
%**************************************************
%
In this section, we study magnetic exchange 
interactions working between quasiparticles.
Consider the magnetic susceptibility to this end.
In the presence of magnetizations, spin and charge 
channels of the susceptibility are coupled 
with each other in general.
In the symmetrical case, however, they are separated 
and the magnetic susceptibility can be expressed as
\begin{equation}
\chi_s(i\omega_l,{\bf q},m)
= \frac{2\pi_s(i\omega_l,{\bf q},m)}
{1 - U \pi_s(i\omega_l,{\bf q},m) },
\label{chis} %----------------------------------------
\end{equation}
with
$\pi_s = \frac1{2} \sum_{\sigma\sigma'} \sigma\sigma'
\pi_{\sigma\sigma'}$,
where $\pi_{\sigma\sigma'}$ is 
the irreducible polarization function.
Similarly,
the magnetic susceptibility of the MAM is given by
\begin{equation}
\tilde{\chi}_s(i\omega_l,m) =
\frac{2 \tilde{\pi}_s(i\omega_l,m)}
{1 - U \tilde{\pi}_s(i\omega_l,m)},
\end{equation}
with $\tilde{\pi}_s$ the single-site polarization
function for the spin channel.
The static and zero-temperature limit of 
$\tilde{\chi}_s$ is approximately given by 
\cite{pre}
\begin{equation}
\lim_{T \to 0}\tilde{\chi}_s(+i0,m) 
= \frac1{k_BT_K} (1-m^2)^{\frac3{2}}.
\end{equation}
%
%====================================================
\begin{figure}
\begin{center}
\epsfile{file=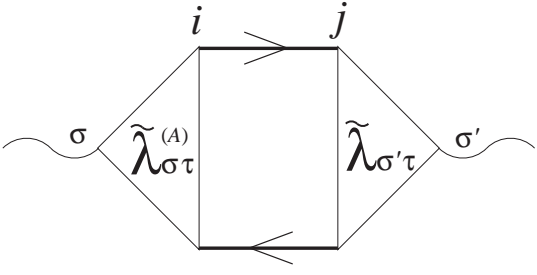,scale=0.45}
\end{center}
\caption[1]{
Diagram for $\pi_{\sigma\sigma'}$
in the site representation.
A thick line with an arrow stands for 
the Green's function.
Overcounted contributions should be subtracted
(see the text).
}
\label{Fig-pi}
\end{figure}
%====================================================
\noindent
Because $\chi_s$ is also of order $1/k_BT_K$,
Eq.~(\ref{chis}) can be rewritten in the form
\begin{equation}
\chi_s(i\omega_l,{\bf q},m) =
\frac{\tilde{\chi}_s(i\omega_l,m)}
{1-\frac1{4}I_s(i\omega_l,{\bf q},m)
\tilde{\chi}_s(i\omega_l,m)},
\label{chis2} %-----------------------------------------
\end{equation}
where
$I_s(i\omega_l,{\bf q},m)$
is the intersite exchange interaction, which is given by
\begin{equation}
I_s(i\omega_l,{\bf q},m) 
= 2 U^2\{ \pi_s(i\omega_l,{\bf q},m)
- \tilde{\pi}_s(i\omega_l,m)\},
\label{Is} %-------------------------------------------
\end{equation}
to leading order in $k_BT_K/U$.
Note that
Eq.~(\ref{chis2}) is consistent with 
the physical picture of the Kondo lattice
that local quantum spin fluctuations at each site are
connected with one another by the intersite interaction.
In order to calculate $I_s$, we consider a two-line
diagram shown in Fig.~\ref{Fig-pi} for the polarization 
part $\pi_{\sigma\sigma'}$, which remains finite 
even in infinite dimensions for specific ${\bf q}$'s.
In other words, the two-line diagram is a leading
term with respect to $1/d$ for specific ${\bf q}$'s, 
with $d$ the spatial dimensionality.
In Fig.~\ref{Fig-pi}, 
$\tilde{\lambda}^{(A)}_{\sigma\tau}$
is a three-point vertex function of the MAM
and $\tilde{\lambda}_{\sigma\tau}$ denotes that of
the periodic Anderson model of $d\to +\infty$ limit.
Note that $\tilde{\lambda}^{(A)}_{\sigma\tau}$
is a single-site term, 
while $\tilde{\lambda}_{\sigma\tau}$ contains 
multi-site terms with respect to both $U$ and $m$.
It follows from Fig.~\ref{Fig-pi} that
\begin{eqnarray}
&& I_s(i\omega_l,{\bf q},m)
 =  - \frac{U^2}{\beta} \sum_{n\sigma}
\tilde{\lambda}^{(A)}_s
(i\varepsilon_n,i\varepsilon_n+i\omega_l,m)
\nonumber \\
&& \mbox{\hspace*{1em}}
\times
\tilde{\lambda}_s
(i\varepsilon_n,i\varepsilon_n+i\omega_l,{\bf q},m)
\nonumber \\
&& \mbox{\hspace*{1em}}
\times
\Bigl\{
\frac1{N}\sum_{\bf k}
G_{ff\sigma}(i\varepsilon_n,{\bf k},m)
G_{ff\sigma}(i\varepsilon_n+i\omega_l,{\bf k}+{\bf q},m)
\nonumber \\
&&  \mbox{\hspace*{3em}}
-
\tilde{G}_{ff\sigma}(i\varepsilon_n,m)
\tilde{G}_{ff\sigma}(i\varepsilon_n+i\omega_l,m)
\Bigr\},
\label{Is2} %-------------------------------------------
\end{eqnarray}
where
$\tilde{\lambda}^{(A)}_s \equiv
\tilde{\lambda}^{(A)}_{\uparrow\uparrow} -
\tilde{\lambda}^{(A)}_{\downarrow\uparrow}$ and
$\tilde{\lambda}_s \equiv
\tilde{\lambda}_{\uparrow\uparrow} -
\tilde{\lambda}_{\downarrow\uparrow}$,
respectively.
In Eq.~(\ref{Is2}), the second term is necessary 
not only to subtract the single-site portion
but also to avoid overcountings.

The main contribution to $I_s$ is divided into two parts:
\cite{exch}
\begin{equation}
I_s(i\omega_l,{\bf q},m)
= J_s({\bf q}) + J_Q(i\omega_l,{\bf q},m).
\end{equation}
The first term is an exchange interaction caused by
the virtual exchange of high-energy spin excitations,
which scarcely depends on $m$.
It is composed of three terms;
the conventional superexchange interaction,
an extended superexchange interaction
and an extended RKKY interaction.
\cite{exch}
Because the property of $J_s$ depends on the whole
band structure,
we treat it as a phenomenological parameter.
The second term, $J_Q(i\omega_l,{\bf q},{m})$,
is due to the virtual exchange of low-energy 
spin excitations within quasiparticle bands:
\begin{eqnarray}
&& 
 J_Q(i\omega_l,{\bf q},m) 
=
- \frac{U^2}{\beta}\sum_{n\sigma}
\tilde{\lambda}^{(A)}_s(+i0,+i0,m)
\nonumber \\
&& \hspace*{1em}
\times
\tilde{\lambda}_s(+i0,+i0,{\bf 0},m)
\nonumber \\
&& \hspace*{1em}
\times
\Bigl\{
\frac1{N}\sum_{\bf k}
G^{(c)}_{ff\sigma}(i\varepsilon_n,{\bf k},m)
G^{(c)}_{ff\sigma}
(i\varepsilon_n + i\omega_l,{\bf k}+{\bf q},m)
\nonumber \\
&&
\mbox{\hspace*{3em}}
-
\tilde{G}^{(c)}_{ff\sigma}(i\varepsilon_n,m)
\tilde{G}^{(c)}_{ff\sigma}(i\varepsilon_n+i\omega_l,m)
\Bigr\},
\label{JQ1} %-----------------------------------------
\end{eqnarray}
where energy dependence and ${\bf q}$ dependence
of the vertex parts have been ignored.
The vertex functions in Eq.(\ref{JQ1}) 
are related to the magnetic parts 
of the corresponding self-energy functions:
\begin{equation}
- \frac{U}{2} \tilde{\lambda}^{(A)}_s(+i0,+i0,m) =
\tilde{\phi}_m 
\frac{\partial \Delta E_A(m)}{\partial m},
\end{equation}
and
\begin{equation}
- \frac{U}{2} \tilde{\lambda}_s(+i0,+i0,{\bf 0},m) =
\tilde{\phi}_m 
\frac{\partial \Delta E(m)}{\partial m},
\end{equation}
respectively.
Therefore we have
\begin{eqnarray}
&&J_Q(i\omega_l,{\bf q},m)
\nonumber \\
&&=
4\ \frac{\partial  \Delta E_A }{\partial m} \cdot
\frac{\partial  \Delta E }{\partial m}
\left\{
P(i\omega_l,{\bf q},m) - \tilde{P}(i\omega_l,m)
\right\},
\label{JQ2} %-------------------------------------------
\end{eqnarray}
with
\begin{eqnarray}
&&P(i\omega_l,{\bf q},m) 
\nonumber \\
&&=
\frac1{N}\sum_{\nu\nu'{\bf k}\sigma}
\frac{f( \xi_{\nu'\sigma}({\bf k}+{\bf q},m) )
-f( \xi_{\nu\sigma}({\bf k},m) ) }
{\xi_{\nu\sigma}({\bf k},m)
-\xi_{\nu'\sigma}({\bf k}+{\bf q},m)
+ i\omega_l},
\label{P} %--------------------------------------------
\end{eqnarray}
where Eq.~(\ref{phi/Z}) has been used and
\begin{eqnarray}
&&\tilde{P}(i\omega_l,m)
\nonumber \\
&&=
- \frac{\tilde{\phi}_m^2}{\beta}
\sum_{n\sigma}
\tilde{G}^{(c)}_{ff\sigma}(i\varepsilon_n,m)
\tilde{G}^{(c)}_{ff\sigma}
(i\varepsilon_n + i\omega_l,m).
\label{P-tilde} %-------------------------------------
\end{eqnarray}

In the following part of this section,
we study the static and uniform component of 
Eq.~(\ref{JQ2}), $J_Q(+i0,{\bf 0},m)$.
It is abbreviated to $J_Q(m,x)$ 
with the argument $x$ explicitly written.
Because excitations between different branches in 
Eq.~(\ref{P}) can be neglected for low-energy phenomena,
we obtain
\begin{equation}
P(+i0,{\bf 0},m) = \sum_{\sigma}
\rho^*_{f}(\sigma\Delta E(m)).
\label{ReP0} %-------------------------------------
\end{equation}
%
%====================================================
\begin{figure}
\begin{center}
\epsfile{file=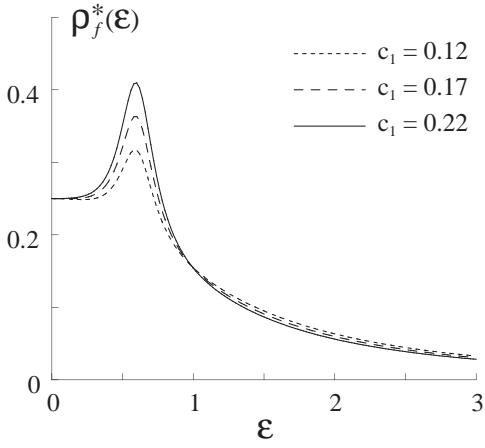,scale=0.40}
\end{center}
\caption[2]{
Phenomenological model for $\rho^*_f(\varepsilon)$
for three values of $c_1$.
Energies are in units of $k_BT_K$.
}
\label{Fig-rho_f}
\end{figure}
%====================================================
\noindent
On the other hand, Eq.~(\ref{P-tilde}) is expanded as
\begin{equation}
\tilde{P}(\omega_+,m)
= \left[\frac{\partial \Delta E_A}{\partial m}
\right]^{-1}   
+ \frac{\pi}{2} i\omega \cdot
\left[\frac{\partial \Delta E_A}{\partial m}
\right]^{-2} + \cdots, 
\label{P-tilde2} %-------------------------------------
\end{equation}
with the use of Eq.~(\ref{sumrule2}),
where $\omega_+ = \omega +i0$.
It can be easily shown 
from Eqs.~(\ref{JQ2}), (\ref{ReP0}) and (\ref{P-tilde2})
that the volume dependence of
$J_Q(m,x)$ is the same as that of $T_K$,
namely, $J_Q(m,x) = e^{-x}J_Q(m,0)$
and the magnitude of $J_Q(m,x)$ is of order $k_BT_K$.
Thus, $J_Q(m,x)$ is scaled with $T_K$,
or the bandwidth of quasiparticles.
This property is responsible for the single-parameter
scaling.
\cite{scaling,pre}
Once an explicit form of $\rho^*_f(\varepsilon)$ is given, 
$J_Q(m,0)$ is straightforwardly 
calculated  as a function of $m$.
As a phenomenological model for $\rho^*_f(\varepsilon)$,
we employ the following model
in the same way as the previous paper, 
\begin{eqnarray}
\rho^*_f(\varepsilon) &=&
\frac1{k_BT_K}\Bigl\{ \frac{c_1}{2}
[D(\bar{\varepsilon};c_2,c_3) 
+ D(\bar{\varepsilon};-c_2,c_3)]
\nonumber \\
&&+ (1 - c_1) D(\bar{\varepsilon};0,c_4) \Bigr\},
\end{eqnarray}
with
$\pi D(y;a,b) = b/[(y-a)^2+b^2]$
and $\bar{\varepsilon} = \varepsilon/k_BT_K$.
In this paper, $c_2=0.60$ and $c_3=0.15$ are assumed,
which are different from previous ones.
Because of the condition $\rho^*_f(0) \simeq 1/4k_BT_K$,
only $c_1$ is variable.
Figure~\ref{Fig-rho_f} represents $\rho^*_f(\varepsilon)$
for three values of $c_1$, and the corresponding results
for $J_Q(m,0)$ are shown in Fig.~\ref{Fig-JQ}.
The exchange interaction $J_Q$
changes from being antiferromagnetic to being 
ferromagnetic with increasing $m$, which is consistent 
with the previous result.
The maximum value of $J_Q$ becomes larger when we raise
the height of the peaks in $\rho^*_f$.
Furthermore, it can be checked by changing $c_2$ 
with $c_1$ fixed that
shifting of the location of the peaks to the band-edge 
side also enhances ferromagnetic instability.
%====================================================
\begin{figure}
\begin{center}
\epsfile{file=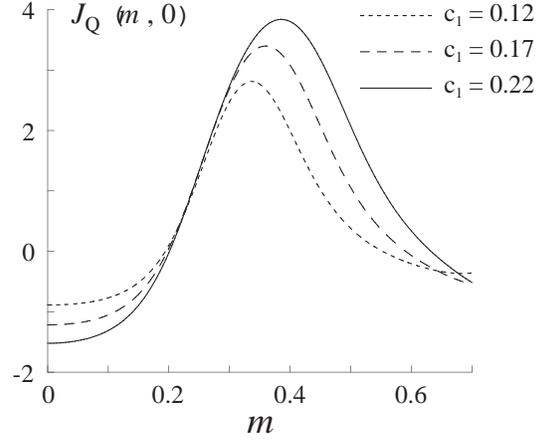,scale=0.40}
\end{center}
\caption[3]{
Exchange interaction $J_Q(m,0)$
(in units of $k_BT_K$)
as a function of $m$.
}
\label{Fig-JQ}
\end{figure}
%====================================================
%
%
%**************************************************
\section{Thermodynamic quantities}
%**************************************************
%
The thermodynamic potential is expressed as 
\begin{equation}
\Omega(m,x) = \Omega_{\rm para}(x) + \tilde{\Omega}(m,x)
+ \Delta\Omega(m,x).
\end{equation}
Here, the first term is that for the paramagnetic state
including a contribution from the lattice system.
It is written as 
\cite{pre}
\begin{equation}
\Omega_{\rm para}(x) = \Omega_{\rm para}(0)
+ k_BT_K(0)\cdot\frac{x^2}{2\kappa},
\end{equation}
where $\kappa$ is a dimensionless compressibility.
The second term and the last term in $\Omega$
are the magnetic single-site and multisite terms; 
they are defined by
$\partial^2 \tilde{\Omega}/\partial m^2 \equiv
1/\tilde{\chi}_s(+i0,m)$,
and
$\partial^2 \Delta\Omega/\partial m^2
\equiv - I_s(+i0,{\bf 0},m)/4$, respectively.
By integrating them, we obtain
\begin{equation}
\tilde{\Omega}(m,x) 
= k_BT_K(x) ( 1 - \sqrt{1 - m^2} \ ),
\end{equation}
and 
\begin{eqnarray}
&&\Delta \Omega(m,x)
\nonumber \\
&& = - \frac{J_s({\bf 0})}{8} m^2
 - \int_0^m dm' \int_0^{m'} dm''
\ \frac{J_Q(m'',x)}{4}.
\end{eqnarray}

\begin{figure}
\begin{center}
\epsfile{file=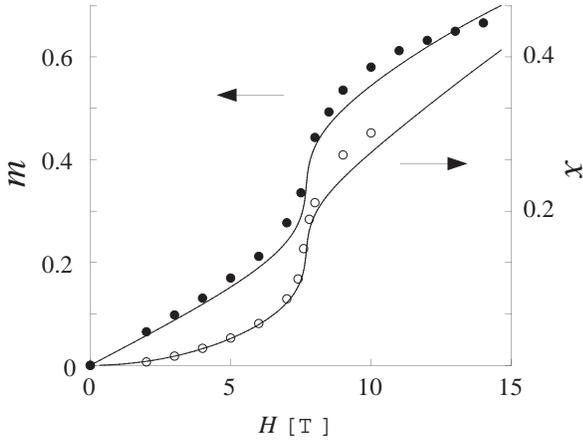,scale=0.49}
\end{center}
\caption[4]{
Magnetization and magnetostriction.
Experimental data are shown by dots and circles,
respectively.
\cite{metamag,striction}
}
\label{Fig-mx}
\end{figure}

In this section,
we confine our study to the case of $c_1 = 0.17$.
The equilibrium values of $m$ and $x$ are determined by
solving the simultaneous equations,
$\partial \Omega/\partial m = H^*$ and
$\partial \Omega/\partial x = 0$;
they are shown in Fig.~\ref{Fig-mx}.
Here, we have set the values of three parameters as
$J_s({\bf 0}) = 0.3 k_BT_K(0)$, $\kappa = 2.4$, and 
$m_0 = 2.0\mu_{\rm B}$
in order to have a better fit with experiments.
The value of $\kappa$ corresponds to 
1.09 [Mbar]$^{-1}$ in the ordinary unit.
It should be mentioned that
if the magnetization process is calculated with
volume fixed constant against the magnetic field,
the metamagnetic crossover is considerably suppressed.
\cite{pre}
%====================================================
%====================================================

\begin{figure}
\begin{center}
\epsfile{file=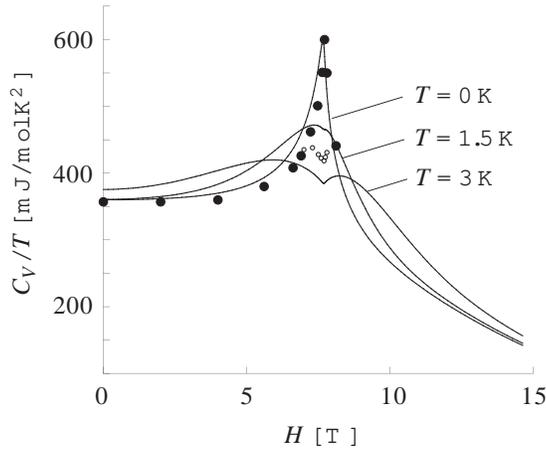,scale=0.45}
\end{center}
\caption[5]{
$C_V/T$ as a function of $H$.
Experimental data at constant pressure for $T=0.25$K 
are shown by dots.
\cite{double-peak}
Circles represent the values of $C_V/T$ at $T=1.5$ K
evaluated from experimental data. 
\cite{Matsuhira}
}
\label{Fig-CV/T-H}
\end{figure}

\begin{figure}
\begin{center}
\epsfile{file=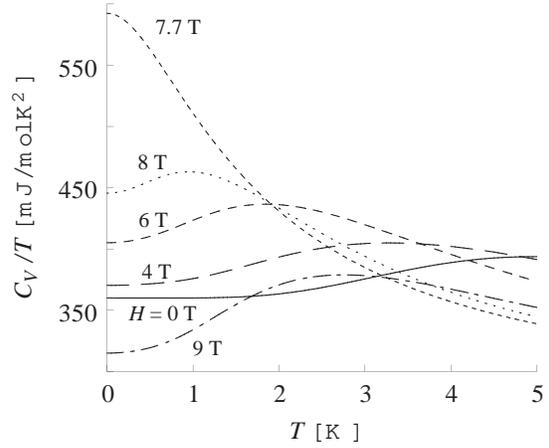,scale=0.45}
\end{center}
\caption[6]{
$C_V/T$ versus $T$
for various values of the magnetic field.
}
\label{Fig-CV/T-T}
\end{figure}

Let us next study the specific heat.
In the presence of magnetizations,
it is given by Eq.(\ref{PAM-heat}) 
with $\rho^*(\varepsilon)$ replaced by
$\rho^*_f(\varepsilon+\sigma\Delta E(m))$.
Figure~\ref{Fig-CV/T-H} shows $C_V/T-H$ curves for three 
values of temperature; $T = 0$ K, $1.5$ K, and $3$ K.
The double-peak structure is qualitatively reproduced
in the present model.
Note that the experiment
\cite{double-peak}
was performed at constant pressure.
From the experimental data,
the specific heat at constant volume has been estimated;
\cite{Matsuhira}
it is also shown in Fig.~\ref{Fig-CV/T-H} for $T = 1.5$ K.
In our calculation,
no double-peak appears at $T = 1.5$ K,
whereas it is observed above $T \simeq 1.0$ K
in the experiment.
A probable reason for this discrepancy is 
the easy estimation of 
the value of $T_K$ based on Eq.~(\ref{phi-TK}).
In addition to that, a contribution from 
ferromagnetic spin fluctuations or paramagnons, 
which is excluded in the SSA, may not be negligible.
Figure~\ref{Fig-CV/T-T} shows the temperature dependence 
of $C_V/T$ for various values of the field,
which is consistent with the experiment
(Fig.~2 of reference 6).
%====================================================
%====================================================
%====================================================
%====================================================

\begin{figure}
\begin{center}
\epsfile{file=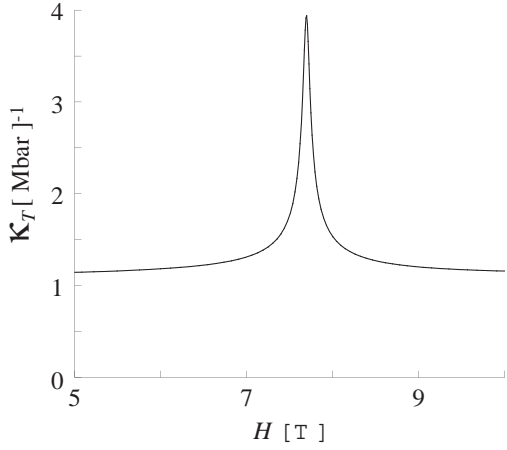,scale=0.45}
\end{center}
\caption[7]{
Isothermal compressibility as a function of $H$.
}
\label{Fig-kappa}
\end{figure}

Finally, the isothermal compressibility is given by
\begin{equation}
\frac1{\kappa_T(H)} = \frac{N\Gamma^2}{V_0}
\left\{
\frac{\partial^2 \Omega}{\partial x^2}
+ \frac{\partial^2 \Omega}{\partial x\partial m}
\cdot \left( \frac{\partial m}{\partial x} \right)_H
\right\}.
\end{equation}
As shown in Fig.~\ref{Fig-kappa},
it is strongly enhanced at $H_M$.
Matsuhira {\it et al} 
\cite{Matsuhira}
have studied the field variation of the compressibility 
by using thermodynamic relations and assuming
the single-parameter scaling.
Our calculation shows a good agreement with their result.
%
%
%**************************************************
\section{Discussion}
%**************************************************
%
It is worth while to comment on spin fluctuations
around the zone center, ${\bf q}\simeq 0$.
The imaginary part of Eq.~(\ref{P}) is expanded as
\begin{equation}
{\rm Im}\ P(\omega_+,{\bf q},m)
= c(m)\cdot\frac{\omega}{q} + \cdots,
\label{ImP} %----------------------------------
\end{equation}
%====================================================
%====================================================
\noindent
for small ${\bf q}$ and small $\omega/|{\bf q}|$, where 
the coefficient $c(m)$ depends on the band structure.
On the other hand, the inverse of the local 
susceptibility is expanded as
\cite{Shiba}
\begin{equation}
\frac1{\tilde{\chi}_s(\omega_+,m)} 
= \frac1{\tilde{\chi}_s(+i0,m)}
-\frac{\pi}{2}i\omega +\cdots.
\label{inverse-chi} %----------------------------------
\end{equation}
Because the $\omega$-linear terms in Eqs.~(\ref{P-tilde2})
and (\ref{inverse-chi}) can be neglected compared with
the $\omega/q$ term in Eq.~(\ref{ImP}) 
for small ${\bf q}$,
\cite{comment3}
the total susceptibility is written in the form
\begin{equation}
\chi_s(\omega_+,{\bf q},m) 
\simeq
\frac{1}
{{\rm Re}\ [1/\chi_s(\omega_+,{\bf q},m)]
-iC(m)\cdot\omega/q}, 
\label{chis3} %----------------------------------
\end{equation}
with $C(m) = \partial \Delta E_A/\partial m \cdot
\partial \Delta E/\partial m \cdot c(m)$.
Since the system is close to ferromagnetic instability 
in the metamagnetic crossover region;
${\rm Re}\ [1/\chi_s(+i0,{\bf 0},m)] \ll k_BT_K$,
the imaginary part of Eq.~(\ref{chis3}) is
expected to be enhanced around the metamagnetic point
for small $\omega$.

On the other hand,
spin fluctuations in the Kondo lattice have been studied
assuming the $\omega$-independent RKKY interaction,
\begin{equation}
\chi_s(\omega_+,{\bf q},m) 
=\frac{\tilde{\chi}_s(\omega_+,m)}
{1 - I_{\rm RKKY}({\bf q})\tilde{\chi}_s(\omega_+,m)},
\label{RKKY} %----------------------------------
\end{equation}
instead of Eq.~(\ref{chis2}) without any justification.
In the low-energy region, however,
Eq.~(\ref{RKKY}) is expressed as
\begin{equation}
\chi_s(\omega_+,{\bf q},m) 
\simeq
\frac{1}
{{\rm Re}\ [1/\chi_s(\omega_+,{\bf q},m)]
%-\frac{\pi}{2}i\omega},
- \pi i\omega/2},
\end{equation}
and behaves quite differently from
Eq.~(\ref{chis3}) around the zone center.
Thus the argument based on Eq.~(\ref{RKKY}) is 
apparently inadequate.
Indeed, recent results of neutron scattering
measurements
\cite{Masugu}
cannot be understood from Eq.~(\ref{RKKY}).
\begin{acknowledgments}
The authors are grateful to
Prof. Toshiro Sakakibara, 
Prof. Ken-ichi Kumagai,
Prof. Takafumi Kita,
and Dr. Kazuyuki Matsuhira
for useful discussions and comments.
This work was partly supported by
a Grant-in-Aid for Scientific Research (C) No. 08640434
from the Ministry of Education, Science, Sports, and
Culture of Japan.
\end{acknowledgments}
%
%
%
%\begin{appendix}
%\end{appendix}
%
%
%

%
%

\begin{references}
\bibitem{comment1}
In the previous paper, we approximately derived
$\tilde{\phi}_m \propto (1-m^2)^{3/2}/\cos^2(\pi m/2)$.
However, its $m$ dependence does not play any essential 
role in the field region of interest.
\bibitem{comment3}
When every leading term in $1/d$ is correctly included,
the $\omega$-linear terms in Eqs.~(\ref{P-tilde2})
and (\ref{inverse-chi}) are canceled with each other.
In the present theory, however,
the cancellation is incomplete except at zero fields
because there is a minor contribution that is not
included in the diagram shown in Fig.~\ref{Fig-pi} 
to leading order in $1/d$.
\end{references}
\end{document}